4*International Conference on Global Optimization and Its Applications 2022 (ICoGOIA 2022), Melaka, November 26-27, 2022*# Optimization of Convolutional Neural Network Hyperparameter for Medical Image Diagnosis using Metaheuristic Algorithms: A short Recent Review (2019-2022)

Qusay Shihab Hamad[1,3], Hussein Samma[2], Shahrel Azmin Suandi[1*]

[1,2]School of Electrical and Electronic Engineering, Engineering Campus, Universiti Sains Malaysia, 14300 Nibong Tebal, Penang, Malaysia
[2]SDAIA-KFUPM Joint Research Center for Artificial Intelligence (JRC-AI) King Fahd University of Petroleum and Minerals, Dhahran, Saudi Arabia
[3]University of Information Technology and Communications (UOITC), Baghdad, Iraq

*Corresponding author email:* shahrel@usm.my
**Abstract**

Convolutional Neural Networks (CNNs) have been successfully utilized in the medical diagnosis of many illnesses. Nevertheless, identifying the optimal architecture and hyperparameters among the available possibilities might be a substantial challenge. Typically, CNN hyperparameter selection is performed manually. Nonetheless, this is a computationally costly procedure, as numerous rounds of hyperparameter settings must be evaluated to determine which produces the best results. Choosing the proper hyperparameter settings has always been a crucial and challenging task, as it depends on the researcher's knowledge and experience. This study will present work done in recent years on the usage of metaheuristic optimization algorithms in the CNN optimization process. It looks at a number of recent studies that focus on the use of optimization methods to optimize hyperparameters in order to find high-performing CNNs. This helps researchers figure out how to set hyperparameters efficiently.

## 1. Introduction

In medical image diagnosis, deep learning recently had a significant role because it had a superior ability to extract features. CNN is a type of DL that has a particular structure. CNN is commonly used in image detection, natural language processing, speech recognition, and many other areas since it uses the convolution operation to derive a collection of features from images (Z. Wang et al., 2021)(Guo, Li, & Zhan, 2021). Due to CNN's robust behavior, many variants have been proposed in recent years. The first model proposed by LeCun et al. (1989) is named LeNet-5. Then AlexNet was proposed by Krizhevsky et al. (2012). Following this, many CNNs were proposed to improve classification accuracies, such as GoogleNet (Szegedy et al., 2015), ResNet (He, Zhang, Ren, & Sun, 2016), DenseNet (Huang, Liu, van der Maaten, & Weinberger, 2016), VGGNet (Simonyan & Zisserman, 2014), etc. Figure 1 illustrates a typical CNN.

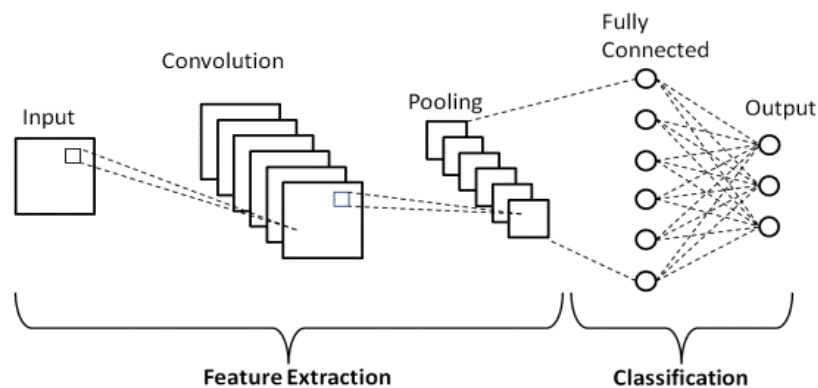

**Figure 1**: CNN example

Considering the various accessible topologies of neural networks, determining the optimal combination of hyperparameters that reduce or maximize the objective function is one of the most challenging aspects of any Deep Learning project (Lacerda, Barros, Albuquerque, & Conci, 2021). Creating a static search space can be a considerable issue when a big number of characteristics are taken into account.

It is tough to select the appropriate values for hyperparameters since it depends not only on one's level of experience but also on her/his ability to acquire knowledge from every round of attempts. Typically, fine-tuning hyperparameters is performed by hand in a time-consuming trial-and-error operation. Many rounds of time-consuming CNN training are needed to test various hyperparameter configurations. New CNNs, on the other hand, tend to have more layers, resulting in an increase in the number of hyperparameters. For example, although AlexNet (Krizhevsky et al., 2012) has only 27 hyperparameters, the successors GoogleNet (Szegedy et al., 2015), ResNet (He et al., 2016), and DenseNet (Huang et al., 2016) have a total of 78, 150, 376 hyperparameters, respectively. As a result, manually pinpointing a close-to-optimal hyperparameter configuration for a CNN at a fair cost is virtually impossible, hindering CNN adaptation for many real-world problems (Y. Wang, Zhang, & Zhang, 2019).

Despite the significance of CNN, as mentioned above, the CNN structure is highly complex. So, how one builds a successful CNN that offers the highest degree of accuracy? For example, how many convolution layers and kernels should be used in each layer? Is the activation mechanism the most effective? What is the worth learning rate? No one has ever been able to give the correct answer to these questions because all CNN hyperparameters are based on two methods: trial and error and historical experiences (He et al., 2016)(Huang et al., 2016). It takes way too much time and effort to build CNN by trial and error. Furthermore, the final model provided by these methods often fails in a locally optimal setting, and this model does not generate the globally best hyperparameters setting. As a result, the most effective way to find the best CNN hyperparameters is to approach the problem as an optimization problem and solve it with robust optimization algorithms (Guo et al., 2021)(Goel, Murugan, Mirjalili, & Chakrabartty, 2022)(Y. Wang et al., 2019)(Fernandes Junior & Yen, 2019)(Kabir Anaraki, Ayati, & Kazemi, 2019)(Darwish, Ezzat, & Hassanien, 2020)(Sun, Xue, Zhang, & Yen, 2020)(Bacanin, Bezdan, Tuba, Strumberger, & Tuba, 2020)(Zatarain Cabada, Rodriguez Rangel, Barron Estrada, & Cardenas Lopez, 2020)(Goel, Murugan, Mirjalili, & Chakrabartty, 2021)(Pathan, Siddalingaswamy, & Ali, 2021)(Zhu et al., 2021) (Gaspar et al., 2021)(Balaha, El-Gendy, & Saafan, 2021)(Deepa & Chokkalingam, 2022)(Gonçalves, Souza, & Fernandes, 2022).

The followings are the main contributions of this study:
1. Provide the reader with a new trend in the application of optimization algorithms to CNN optimization.
2. Elucidate the hyperparameters that were chosen in recent researches.

The following is the organization of this paper: The basic concept of CNN hyperparameter is discussed in Section 2, and an overview of optimization algorithms is presented in Section 3. Section 4 clarifies the methodology of using an optimization algorithm to optimize hyperparameters. The researchers used optimization algorithms to optimize hyperparameter, which are explained in Section 5. The research discussion is presented in Section 6. The conclusion and possible research areas are addressed in the last part, Section 7.

## 2. The Basic Concept of CNN Hyperparameter

Along with the parameters that should be trained, such as weight and bias, there are certain additional parameters called hyperparameters that must be established prior to the network training (Guo et al., 2021). CNN's hyperparameters include how many convolutional layers (CL)? The number and size of CL kernels; the kind of pooling; the type of activation functions used in different layers; how many neurons are in the fully connected (FC) layers? The dropout rate, the maximum epoch, the learning rate, etc.

## 3. Metaheuristic Optimization Algorithms

Optimization algorithms are grouped into two fundamental sections: population-based and individual-based. In the population-based algorithm, a series of random solutions are generated as the starting point. This set is then iteratively improved with many operators until an end condition is satisfied. According to the literature, population-based algorithms are more often used than individual-based algorithms across a wide variety of areas. This is because such algorithms are more adept at avoiding local optima (Mirjalili, Dong, Lewis, & Reviews, 2020)(Abualigah, Diabat, Mirjalili, Abd Elaziz, & Gandomi, 2021)(Hamad, Samma, Suandi, & Saleh, 2022)(Hamad, Samma, Suandi, & Mohamad-Saleh, 2022).

## 4. Methodology

The main concept of using an optimization algorithm to find the optimal hyperparameters resulted in improved CNN results. This process can be divided into two main phases: the training phase and the testing phase as illustrated in Figure 2. Through the training phase, the search agents of the optimization algorithm try to find the optimal value for the CNN hyperparameter, this process is done by each agent selecting random numbers for the hyperparameter and then using training data to train the CNN and evaluate its performance the result of the performance considered by optimization algorithm to evaluate its agents, the optimization algorithm saves the best agent's value (setting of CNN that got the highest performance) this process repeated for each search agents at each iteration let say optimization algorithm has 30 search agent and a maximum number of iterations is set to 10, in total the process of searching for optimal hyperparameters will be repeated 300 times. At the last iteration, the optimization algorithm based on its self-evaluation for all its search agents will return back the optimal hyperparameter selected through the training phase. The selected parameters are evaluated through a new dataset in the testing phase.

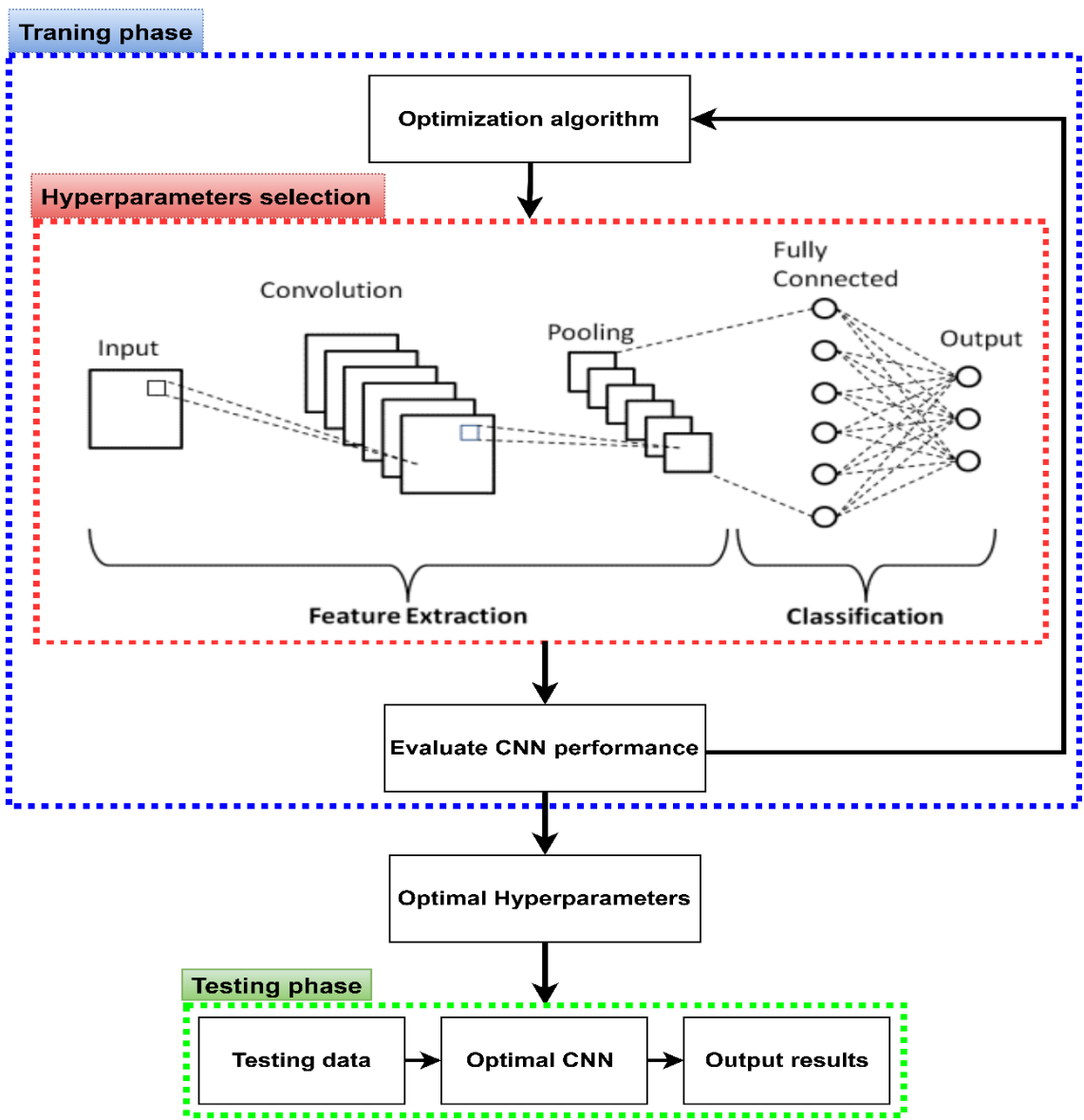

**Figure 2**: The CNN hyperparameters optimization process

## 5. Optimize CNN Hyperparameter Through Metaheuristic Optimization Algorithms

This section introduces a collection of research efforts aimed at resolving the problem of determining the optimal CNN hyperparameters through the use of metaheuristic optimization algorithms. A non-invasive glioma classification method using magnetic resonance imaging (MRI) that combined Genetic algorithms (GA) and CNNs was suggested by Anaraki et al. (2019). The CNN structure was evolved using GA, avoiding the traditional trial-and-error method of selecting the best CNN architecture for a given problem. Two case studies were used to test the proposed procedure. In the first case study, 90.9 % classification accuracy was achieved for three different glioma classes. Glioma, meningioma, and pituitary tumor types were identified with 94.2 % precision in the second study. Polsinelli et al. (2020) proposed a light CNN for detecting COVID-19 from computer tomography (CT) imaging of the chest. They used Bayesian optimization (BO) to select the optimal hyperparameters for SqueezeNet. They selected three hyperparameters in their work: initial learning rate, momentum, and L2-Regularization. The proposed model achieved an accuracy of 85.03%.

Goel et al. (2021) used the Grey Wolf Optimizer (GWO) algorithm to optimize the CNN's hyperparameters for classifying COVID-19, normal, and pneumonia patients from chest X-ray images. This model is called Optimized Convolutional Neural Network (OptCoNet). The dataset used in the experiments was gathered from various sources; the total number of images is 2700, with 900 for each of COVID-19, Pneumonia, and Normal. The reported classification accuracy of the proposed model was 97.78%. An optimized CNN model was proposed by Pathan et al. (2021) to create an automated image analysis module to identify COVID-19 affected chest X-ray scans. The hyperparameters of CNN were tuned using two optimization algorithms: GWO and Whale Optimization + BAT algorithm (WOA-BAT). The classification results on the two datasets containing Dataset1 (2905 images) and Dataset2 (6342 images). The number of COVID-19 images is (Dataset 1 has 219 images), (Dataset 2 has 576 images). For training and testing, the data was divided into an 80:20 ratio. The hyperparameters that the optimization algorithms managed to optimize are initial learning rate, L2 regularization, Max epochs, Gradient decay factor, and Validation Frequency. Dataset 1 was used for hyperparameter tuning, and the same parameters were used for training and testing Dataset 2. For Dataset 1 and Dataset 2, classification accuracy was reported to be 98.8% and 96.0%, respectively. Balaha et al. (2021) developed a technique for COVID-19 detection in chest computed tomography (CT) images. To optimize the CNN hyperparameters, they used Harris Hawks Optimization (HHO). They compared nine pre-trained CNNs in this study. These models were ResNet50, ResNet101, VGG16, VGG19, Xception, MobileNetV1, MobileNetV2, DenseNet121, and DenseNet169. HHO was tasked with the responsibility of determining the optimal hyperparameters. The dropout ratio, learning ratio, and batch size were all used as hyperparameters. The VGG-19 reported the best value using the Stochastic Gradient Descent (SGD) parameters optimizer, 32 batch size, 56% dropout ratio, and 80% learning ratio. Goel et al. (2022) proposed a new model for COVID-19 detection called Multi-COVID-Net, in which the CNN's hyperparameters were optimized using the Multi-Objective Grasshopper Optimization Algorithm (MOGOA). Inceptionv3 and ResNet50 were the two CNNs used. After optimizing the hyperparameters of both networks for the COVID-19, normal, and pneumonia datasets, features from both networks are extracted. These features are combined and fed into SVM for classification into the correct label. The batch size, learning rate, momentum, number of epochs, and regularization coefficient are the hyperparameters of the CNN that were chosen in this study. The dataset contains 2700 images, 900 for each class, and is divided into 70% for training and 30% for testing. The proposed model's accuracy was determined to be 98.21%. Deepa and Chokkalingam (2022) optimized the learning rate and batch size of VGG16 using the Arithmetic Optimization Algorithm (AOA). The case study for this research was the detection of Alzheimer's disease using MRI images. The reported results demonstrated that AOA increases the accuracy of VGG16 classification. A model for detecting COVID-19 from chest X-ray images has been proposed by Jalali et al. (2022). In this study, they developed a competitive swarm optimizer (CSO), a robust and effective PSO variation, to attain the best CNN hyperparameters and boost the classification accuracy of X-ray images. In the suggested model, the MCSO will optimize eleven CNN hyperparameters. These hyperparameters are the number of CL, number of filters in CL, filter size, activation function type, dropout rate, max-pooling size, learning rate, momentum rate, optimizer type, number of epochs, and batch size. In this work, KNN is used as a classifier. Ma et al. (2022) used PSO to propose a CNN to detect cervical tumors. In this research, PSO works on selecting the best hyperparameters like the number of convolution layers, the number of filters in convolution, filter size, the type of pooling layer, and selecting the number of neurons in the fully connected layer. In the proposed model, the last layer should be a fully connected layer. Gonçalves et al. (2022) used two optimization algorithms, GA and PSO, to optimize several hyperparameters and the architecture of fully connected layers from three CNNs (VGG-16, ResNet-50, and DenseNet-201). Breast cancer detection was the case study. The following hyperparameters were used in this study: The dropout rate and after which layers there is dropout, number of FC, number of neurons in each FC, and learning rate. The results that were published show that both proposed optimizations did better than most manual hyperparameter tests, but in experiments, the GA did better than the PSO. Table 1 summarizes all efforts made by researchers to optimize hyperparameters.

Table 1: Summary of Research on Hyperparameter Optimization

| Reference | Model Name | Metaheuristic Algorithm | Method Type | Hyperparameters optimized | Problem Type |
|---|---|---|---|---|---|
| Kabir Anaraki et al. (2019) | - | GA | CNN from Scratch | number of CL and FC, Type of pooling layers, after | Classify Glioma brain tumor |
| | | | | which layers there is dropout, number and size of filter, activation functions, optimizer type, learning rate, and dropout rate. | using MRI images |
| Polsinelli et al. (2020) | CNN-2 | BO | SqueezeNet | Learning Rate, Momentum, L2-Regularization. | Detect COVID-19 patients from chest CT images |
| Goel et al. (2021) | OptCoNet | GWO | CNN from Scratch | momentum, learning rate, maximum epoch, validation frequency, and L2 regularization. | Detect COVID-19 patients from chest X-ray images. |
| Pathan et al. (2021) | GWO-CNN | GWO | CNN from Scratch | learning rate, L2 regularization, maximum epochs, gradient decay factor, and validation frequency. | Detect COVID-19 patients from chest X-ray images. |
| Pathan et al. (2021) | WOA-BAT-CNN | WOA-BAT | CNN from Scratch | learning rate, L2 regularization, maximum epochs, gradient decay factor, and validation frequency. | Detect COVID-19 patients from chest X-ray images. |
| Balaha et al. (2021) | CovH2SD | HHO | VGG19 | dropout rate, learning rate, and batch size | Detect COVID-19 patients from chest CT images |
| Goel et al. (2022) | Multi-COVID-Net | MOGOA | Inceptionv3 and ResNet50 | batch size, learning rate, momentum, maximum epochs, and L2-Regularization | Detect COVID-19 patients from chest X-ray images. |
| Deepa and Chokkalingam (2022) | - | AOA | VGG16 | learning rate and batch size | detection of Alzheimer's disease using MRI images |
| Jalali et al.(Jalali et al., 2022) | MCSO-CNN | MCSO | CNN from Scratch | Number of CL, number of filters in CL, filter size, activation function type, dropout rate, max-pooling size, learning rate, momentum rate, optimizer type, number of epochs, and batch size. | Detect COVID-19 patients from chest X-ray images. |
| Ma et al.(2022) | PSO-CNN | PSO | CNN from Scratch | Number of CL, number of filters in CL, filter size, type of pooling layer, number FC, number of neurons of FC | Cervical tumors |
| Gonçalves et al. (2022) | - - | GA PSO | VGG16, ResNet-50, and DenseNet-201 | learning rate, number of FC, number of neurons in each FC, after which layers there is dropout, and dropout rate | Breast cancer detection |

Table 2 clarifies the types of hyperparameters and how many times they are used in the literature. The learning rate is located in the first rank with 9 times, while batch size, maximum epoch, and L2 regularization and maximum epoch are located in the second rank with 5 times, and max-pooling size is located in the last rank with one time only.

Table 2: Summary of how many times hyperparameter used in literature

| Hyperparameter | Number of CL | Number and size of filter | Type of pooling layers | Number of FC | Number of neurons in FC | Activation functions |
|---|---|---|---|---|---|---|
| How many times used | 3 | 3 | 2 | 3 | 2 | 2 |
| **Hyperparameter** | **batch size** | **Momentum** | **L2-Regularization** | **Dropout rate** | Maximum epoch | Validation frequency |
| How many times used | 4 | 4 | 5 | 4 | 5 | 3 |
| **Hyperparameter** | After which layers there is dropout | Learning rate | Optimizer type | Gradient decay factor | Max-pooling size | |
| How many times used | 2 | 9 | 2 | 2 | 1 | |

## 6. Discussion

This section expands on the research presented in the previous section. Researchers chose various types of hyperparameters to try to improve the CNN model's accuracy in various types of classification problems. Some researchers choose a large number of hyperparameters to achieve high performance, but these experiments take a long time and require an efficient machine to complete in a reasonable amount of time. Some hyperparameters are frequently used because they have a large impact on model performance. Using pre-trained CNN also saves time and effort because there is no need to work with a large number of parameters; only a few parameters, such as the learning rate, will suffice.

## 7. Conclusion and future work

This paper presented a recent review of the improvements to CNN accuracy made possible by the use of optimization algorithms to assist in automatically setting hyperparameters. Optimization algorithms will seek optimal hyperparameter values that reduce effort and prevent ineffective CNNs. In this paper, we were able to see how researchers employed various optimization algorithms to configure a vast array of hyperparameters. Each of these architectures aims to address a problem associated with a particular medical image diagnosis. More than half of the proposed research uses optimization algorithms to find optimal CNN hyperparameters from scratch. 63.63% of the researches were used to solve the problem of COVID-19 detection because it is a new pandemic, and the results indicate that optimizing CNN through an optimization algorithm can provide a good result for COVID-19 detection. This paper intends to draw the reader's attention to a very important area of research, and it is possible to use alternative optimization algorithms to attempt to improve various CNN types.

In future work, we intend to conduct a comprehensive survey of all types of methods used to propose CNN that are employed in various types of applications and attempt to identify the best approach that researchers can use to propose a CNN with the highest performance.

## Acknowledgments

This research is supported by the Malaysia Ministry of Higher Education (MOHE) Fundamental Research Grant Scheme (FRGS), no. FRGS/1/2019/ICT02/USM/03/3.